\documentclass[12pt,preprint]{aastex}
\usepackage{epsfig}
\begin{document}
\newcommand{\spi}{{\it Spitzer}}
\newcommand{\molh}{H$_2$}
\newcommand{\napjs}{{\it Astrophys. J. Supp.}}
\newcommand{\napj}{{\it Astrophys. J.}}
\newcommand{\naj}{{\it Astron. J.}}
\newcommand{\etal}{{\it et al.}}
\newcommand{\naanda}{{\it Astron. Astrophys.}}
\newcommand{\nmnras}{{\it Mon. Not. R. Astron. Soc.}}

\title{\molh\ emission arises outside photodissociation regions in ultra-luminous infrared galaxies}

\author{Nadia L. Zakamska\\
Institute for Advanced Study, Einstein Dr., Princeton, NJ 08540}

{\sl Ultra-luminous infrared galaxies are among the most luminous objects in the local universe and are thought to be powered by intense star formation$^{1,2}$. It has been shown that in these objects the rotational spectral lines of molecular hydrogen observed at mid-infrared wavelengths are not affected by dust obscuration$^3$, leaving unresolved the source of excitation of this emission. Here I report an analysis of archival \spi\ Space Telescope data on ultra-luminous infrared galaxies and demonstrate that star formation regions are buried inside optically thick clouds of gas and dust, so that dust obscuration affects star-formation indicators but not molecular hydrogen. I thereby establish that the emission of \molh\ is not co-spatial with the buried starburst activity and originates outside the obscured regions. This is rather surprising in light of the standard view that \molh\ emission is directly associated with star-formation activity$^{3,4,5}$. Instead, I propose that \molh\ emission in these objects traces shocks in the surrounding material, which are in turn excited by interactions with nearby galaxies, and that powerful large-scale shocks cooling by means of \molh\ emission may be much more common than previously thought. In the early universe, a boost in \molh\ emission by this process may speed up the cooling of matter as it collapsed to form the first stars and galaxies and would make these first structures more readily observable$^6$.}

The analysis is based on \spi\ Space Telescope Infrared (IR) Spectrograph$^7$ mid-IR (5$-$35\micron) spectra of 48 ultra-luminous infrared galaxies$^8$ (ULIRGs) publicly available through \spi\ Archive. The objects were selected from the {\it IRAS} 1 Jy sample to include only those ULIRGs that did not have signatures of powerful supermassive black hole activity in their optical spectra$^8$. For these objects I measure the strength of the 10\micron\ silicate absorption feature ($S$[9.7\micron]), the strength of 6\micron\ water ice absorption ($S$[6.0\micron]), the fluxes of \molh\ emission lines, the fluxes of polycyclic aromatic hydrocarbon (PAH) emission features -- which are used as star-formation indicators$^9$ -- and the bolometric fluxes F(IR) of the galaxies (as described in Supplementary Information). In addition, I use other samples of nearby ULIRGs for which some of these measurements are available in the literature$^{3,10,11}$. As a comparison data set, I use the publicly available spectroscopic data from the \spi\ Infrared Nearby Galaxy Survey (SINGS$^{12}$), which I analyze using the same methods as ULIRG spectra. Nuclei of SINGS galaxies that show activity associated with accretion onto a supermassive black hole have been excluded from the comparison sample. Therefore, star formation is the likely dominant energy source for both ULIRGs and comparison galaxies selected for this Letter.

\begin{figure}
\epsscale{0.5}
\plotone{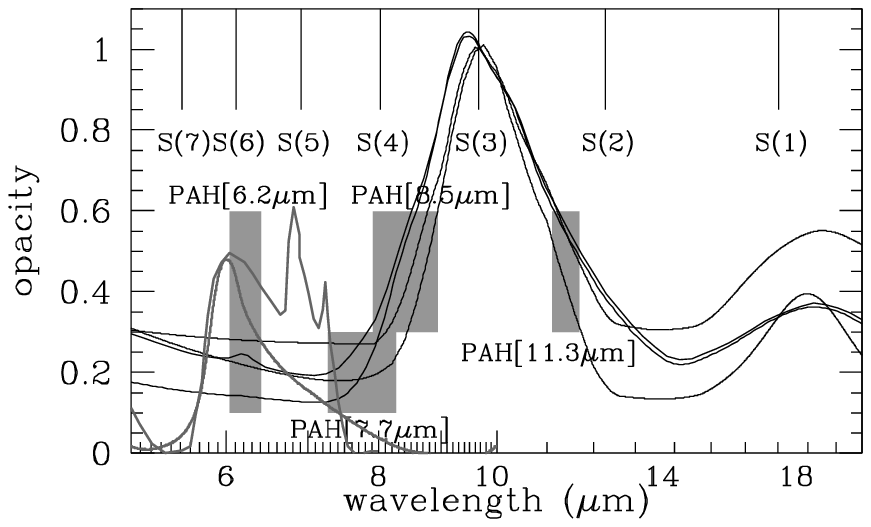}
\figcaption{{\bf Wavelengths of emission features present in ULIRG spectra and representative opacity curves.} If emission regions are embedded in dust or ice, those emission features that are near the peaks of opacity should be the ones most strongly affected. If \molh\ emission originates inside silicate dust obscuration (example opacities$^{25,26,27}$ in black), then S(3) should be strongly extincted, S(1) somewhat less so, and other lines less still. If PAH emission originates inside obscuration, the features centered at 8.5\micron\ and 11.3\micron\ should be more affected by silicates than the other ones, while PAH[6.2\micron] is the only feature that may be affected by water ice (grey; smoothed data for the whole complex$^{28}$ and laboratory data for water ice$^{29}$, shown at the maximum strength according to the $S$[6.0\micron]$=0.6S$[9.7\micron] relation -- see Supplementary Information). \label{pic_extinction}}
\end{figure}

Even though the starbursts supply most of the energy output of ULIRGs, which is in excess of $10^{12}$ solar luminosities, only a small fraction of each ULIRG's bolometric luminosity is observed as UV and optical continuum of young stars$^{1}$, as most of this emission is absorbed by dust and re-emitted thermally at IR wavelengths. If the observed PAH and \molh\ emission is powered by the embedded starbursts, then these features should also be extincted by the intervening matter. Because the amount of extinction depends on wavelength, different features will be affected by different amounts (Figure \ref{pic_extinction}); the ones that fall within the strong opacity peak from astronomical silicates at 9.7\micron\ should be affected most strongly. By taking ratios of emission line fluxes within and outside the absorption feature one can test where the emitting region is located relative to the source of opacity. Since ULIRGs are optically thick even at mid-IR wavelengths, with a median silicate absorption strength of 1.6, the effect of opacity is expected to be strong. 

\begin{figure}
\epsscale{1.0}
\plotone{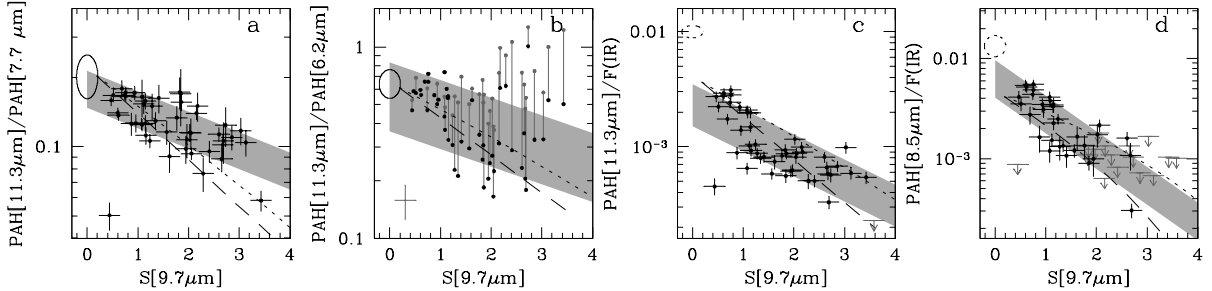}
\figcaption{{\bf PAH features are affected by dust extinction.} This conclusion is evidenced by correlations of the observed ratios of PAH fluxes (a,b) and the PAH/F(IR) ratios (c,d) with the apparent strength of the silicate opacity feature. Spearman rank probabilities of the null hypothesis that the plotted values are uncorrelated are (a) P[NH]$=10^{-4}$, (c) P[NH]$=10^{-5}$, (d) P[NH]$=10^{-5}$. In panel (b), the observed PAH[11.3\micron]/PAH[6.2\micron] ratio (grey circles) is uncorrelated with the absorption strength (P[NH]$=0.22$), but when PAH[6.2\micron] is corrected for water ice absorption (black circles), the correlation becomes apparent (P[NH]$=2\times 10^{-3}$), suggesting that the PAH-emitting region is located behind both silicate absorption and water ice absorption. In each panel (a) and (b), PAH ratios calculated for the comparison sample of nearby star-forming galaxies$^{12,30}$ are shown by an ellipse whose semi-axes are determined by the standard deviation of the corresponding measure. Although PAH ratios are known to vary as a function of physical conditions$^{30}$, PAH ratios seen in ULIRGs are consistent with those found in the comparison sample when extrapolated to low absorption. Dashed and dotted lines illustrate how unobscured ratios would change in the presence of an increasing amount of cold dust between the emitter and the observer for representative opacity curves$^{26,27}$. In panels (c) and (d), the model calculation assumes that all absorbed flux is re-emitted at longer wavelengths, but the total flux does not change. PAHs constitute a higher fraction of the total luminosity output in low-luminosity galaxies (estimates shown with dotted ellipses) than they do in ULIRGs$^{13}$. The grey shaded areas show the 1$\sigma$ range in the vertical offset around the best linear fit for ULIRG data. Filled circles denote detections, arrows denote 3$\sigma$ upper limits, and 1$\sigma$ standard errors are shown in each panel except (b) where individual error bars are omitted for clarity and the median error is shown in the bottom left corner. \label{pic_pah}}
\end{figure}

Indeed, the observed PAH ratios depend on the strength of absorption (Figure \ref{pic_pah}), confirming that the PAH-emitting regions are located behind silicates and water ices. This observation provides a quantitative measure of absorption correction that needs to be applied to PAHs to derive star-formation rates in ULIRGs. The slopes of the observed correlations between PAH emission ratios and strengths and $S$[9.7\micron] are somewhat flatter than those expected if there is a screen of dust between the emission regions and the observer; that is, PAH emission is less absorbed than expected from the simplest model. This difference suggests that the PAH-emitting regions are spread out within the absorbing medium, rather than being concentrated in one central point. Because of the extinction, we are more likely to observe those of the PAH-emitting regions that are closest to us and are therefore the least obscured. 

\begin{figure}
\epsscale{1.0}
\plotone{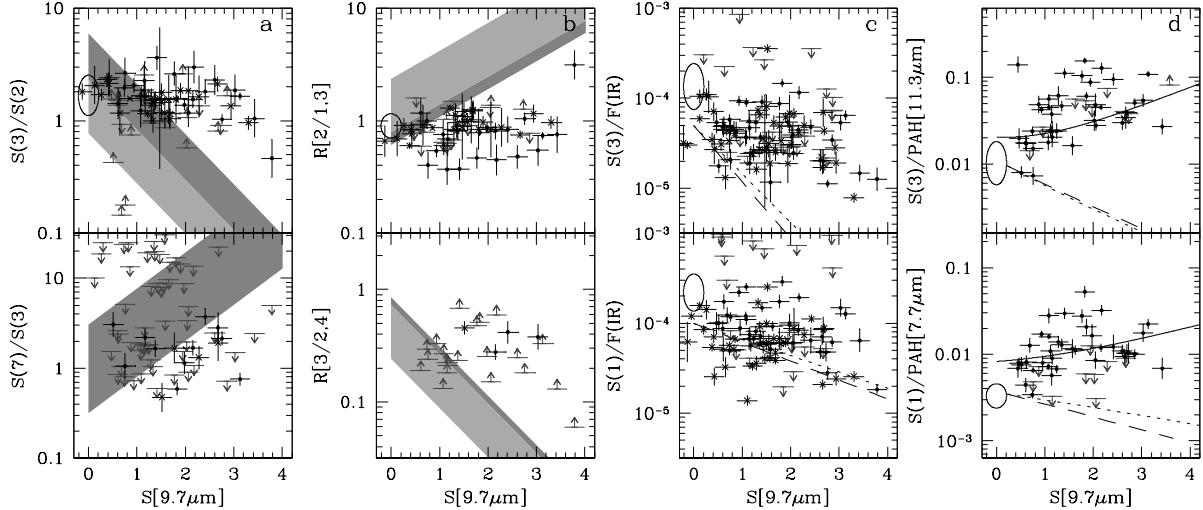}
\figcaption{{\bf \molh\ emission in ULIRGs is not affected by extinction and shows excess over the \molh/PAH ratio observed in normal galaxies.} The ratios of \molh\ fluxes (a,b) and the \molh/F(IR) ratios (c) for ULIRGs from two samples (circles$^8$, crosses$^3$) are uncorrelated with the apparent strength of the silicate absorption. In (b), R[$J_1/J_2,J_3$] is the ratio of the observed flux of the line S($J_1$) to that expected based on S($J_2$) and S($J_3$) assuming all three lines come from a region with a single excitation temperature. Grey areas in (a, b) and lines in (c) show the expected trends of line ratios with apparent silicate strength if \molh\ emission is behind a screen of dust. Models in dark grey assume an ortho-to-para ratio of 3 and a realistic range of excitation temperatures, whereas light grey areas are an extension of the model to include ortho-to-para ratios between 1 and 3. Although correlations are expected if \molh\ is affected by silicate absorption, none are detected (P[NH]$=5-75\%$). In (d), if \molh\ and PAH emission had the same spatial distribution, the \molh/PAH ratios would be expected to decrease (dotted and dashed lines) because dust opacity at the wavelength of the \molh\ line is greater than that at the wavelelngth of the PAH feature, but in fact an increase is observed (P[NH]=0.007, 0.016 for top and bottom). The \molh/PAH ratios are described better by a model that combines an obscured component of \molh\ associated with star formation and an unobscured \molh\ component with the luminosity equal or somewhat greater than that of the buried component (solid lines in panels (d), top and bottom assume $L[{\rm H_2, outer}]/L[{\rm H_2, inner}]=1$ and 1.5, respectively). Arrows denote 3$\sigma$ upper and lower limits, and 1$\sigma$ standard errors are shown for all measurements. Ellipses show flux ratios for comparison galaxies (semi-axes are determined by the standard deviation of the corresponding measure). \label{pic_h2}}
\end{figure}

On the contrary, as was previously concluded$^3$, \molh\ is not affected by absorption. \molh\ fluxes and ratios fail to show any correlation with the strength of silicate opacity (Figure \ref{pic_h2}a-c), even though the S(3) line coincides in wavelength with the peak of the silicate dust opacity curve (Figure \ref{pic_extinction}) and should be very strongly affected. In the absence of an observable dependence, the question arises just how strong a correlation would be expected if \molh\ emission were behind the absorbing material. Molecular hydrogen is known to be present at a range of excitation temperatures$^3$ and ortho-to-para ratios$^4$, so even a simple model should take into account a wide range of physical conditions. To calculate the expected \molh\ line ratios, I assume that ${\rm d}M\propto T^{-p} {\rm d}T$, where ${\rm d}M$ is the mass of \molh\ gas with excitation temperatures in the range $T,T+{\rm d}T$. For each $p$, I calculate the emitted \molh\ spectrum for ortho-to-para ratios between 1 and 3. The range $2.5<p<5.0$ is sufficient to reproduce the observed range of \molh\ excitation diagrams of galaxies (as shown in Supplementary Information). I then calculate \molh\ line ratios expected if the \molh-emitting region were behind a screen of cold dust and select those of the line ratios that are expected to be most sensitive to extinction. 

Even given a generous range of model conditions and intrinsic range of \molh\ line fluxes and ratios, their dependence on extinction would have been unambiguous if it were present in Figures \ref{pic_h2}(a-c). It is also noteworthy that models at zero extinction succeed in reproducing the observed range of line ratios, which confirms that the models adequately represent the range of physical conditions in real objects. Using different line combinations, more than a dozen line ratios can be computed; none shows dependence on $S$[9.7\micron]. Neither the line ratios associated with the cooler component (median excitation temperature $T_{\rm exc}=330$K as measured from  S(1), S(2) and S(3)) nor the ones from the warmer component (median $T_{\rm exc}=1200$K as measured from S(3), S(4) and S(7)) show any evidence for extinction.

The fractional contribution of \molh\ and that of PAHs (extrapolated to low extinction) to the total luminosity is smaller in ULIRGs than in comparison galaxies. For PAHs, this finding is in agreement with the known non-linear relationship between bolometric and PAH luminosities$^{13}$. As for \molh, there has to be a component directly associated with star formation, just like in normal galaxies. However, as the extinction is increased, this component is quickly extinguished, and we can only see \molh\ emission that originates outside obscuration. The luminosity of this component may be estimated by observing that the \molh/PAH ratios in ULIRGs exceed those seen in normal galaxies (Figure \ref{pic_h2}d). The data are well described by a model in which about as much (and up to 50\% more) \molh\ emission is emitted outside the bulk of the obscuration as is directly associated with the starbursts, if the intrinstic PAH/\molh\ ratios are the same in ULIRGs and in comparison galaxies. 

If dust is mixed in with emission regions or has patchy distribution, the correlations with silicate strength would be expected to be less steep than in the `screen of dust' case. If dust is not completely cold, its thermal emission would fill in somewhat the absorption feature, making the observed $S$[9.7\micron] an underestimate of the true optical depth, and the correlations would be expected to be steeper than in `screen of dust' case. However, such effects cannot change the sign of the expected correlations. Since \molh/PAH ratios increase with opacity when a decrease is expected from the models (Figure \ref{pic_h2}d), any model with the same spatial distribution of \molh\ and PAHs is ruled out by the data.  

The first step in understanding the origin of the rotational \molh\ emission in ULIRGs is the determination of the spatial extent of this emission, and in particular, whether this emission is confined to the outer parts of starburst regions or is extended on galaxy-wide scales. One way to proceed is to study the spatial distribution of the ro-vibrational lines in the near-IR, where observational capabilities for spatially resolved spectroscopy or narrow-band imaging are not as limited as in mid-IR. This approach yielded two successful observations of ULIRGs, one of NGC 6240$^{14}$ and one of Arp 220$^{15}$. In both cases the objects are composed of two well-separated merging components, and the near-IR \molh\ emission peaks between them, near the collision front. 

Since most ULIRGs are either actively merging or dynamically disturbed$^{16}$, these observations suggest the possibility that \molh\ lines trace shocks excited by inter-galactic interactions$^{17}$. Recent observations with \spi\ Space Telescope uncovered a class of unusually \molh-luminous extragalactic objects$^{18,19,20}$, in which little or no star-formation is seen. One of these objects, a 40 kiloparsec inter-galactic shock in Stephan's Quintet$^{18,21}$, emits tens of percent of its bolometric luminosity in rotational \molh\ lines. A similar process -- albeit producing twenty times more mid-IR \molh\ luminosity in a median ULIRG than in Stephan's Quintet -- may be operating in the outer parts of ULIRGs. Further support for this hypothesis is provided by observations of optical line emission from six ULIRGs$^{22}$, including four from this Letter, in which a contribution from inter-galactic shocks is directly seen. The median excitation temperature measured between the S(1) and S(3) transitions for ULIRGs with $S$[9.7\micron]$>1$ in which the unobscured component dominates the observed \molh\ flux is 306 K, similar to that measured in Stephan's Quintet between the same transitions (350 K$^{21}$).  

Processes other than inter-galactic shocks could in principle lead to excitation of \molh\ emission. However, as discussed in Supplementary Information, excitation by X-rays, by cosmic rays or by emission due to accretion onto the supermassive black hole are unlikely to be dominant producers of the observed \molh\ emission on energy grounds. Alternatively, \molh\ could also be excited in shocks other than those produced by galactic collisions. For example, it could be due to super-galactic winds or supernova remnants which, having cleared up the environment around them, may be not as obscured as the photon-dominated regions that produce most of the PAH emission. These processes are neither supported nor ruled out as the dominant source of \molh\ excitation by the available data on ULIRGs. If they turn out to be important, this will be in contrast to lower-luminosity starburst galaxies in which these processes do not dominate \molh\ excitation$^{23}$. Future observations of spatial distribution and kinematics of \molh\ emission will help distinguish between intergalactic shocks and other scenarios. 

The observations of ULIRGs presented here raise the possibility that excitation of \molh\ in shocks and their subsequent cooling through the mid-IR rotational lines of \molh\ is a much more common phenomenon than previously thought. Instead of being an astronomical curiosity illustrated by a handful of \molh-luminous objects and a few interacting galaxies, such cooling may be occurring in a wide range of conditions, as the large sample of ULIRGs demonstrates. If similar conditions are commonly encountered in the early universe$^{24}$, when \molh\ is one of the primary coolants, cooling via \molh\ emission in shocks can have a dramatic effect on formation of the first objects. 

{\bf Supplementary Information} is linked to the online version of the paper at \\
www.nature.com/nature.

{\bf Acknowledgments} This work was supported by the \spi\ Space Telescope Fellowship provided by NASA through a contract issued by the Jet Propulsion Laboratory, California Institute of Technology, by the John N. Bahcall fellowship at the Institute for Advanced Study and by the NSF grant AST-0807444. I would like to thank M. Imanishi for providing reduced, flux-calibrated ULIRG data in electronic form and for his permission to use these data for a study of \molh\ emission. I would like to thank P. Goldreich, J. Krolik and S. Davis for discussions, and H. Spoon and L. Hao for providing electronic data.

{\bf Author Information} Reprints and permissions information is available at \\
www.nature.com/reprints. Correspondence and requests for materials should be addressed to N.L.Z. (zakamska@ias.edu). 

\clearpage

{\bf Supplementary Information}

ULIRGs have luminosities $L>10^{12}L_{\odot}$, with most of their emission coming out at infrared wavelengths$^1$. The energy source of these objects remains a matter of controversy. Although the conventional interpretation is that they are powered by starbursts$^2$, some authors argue that buried active galactic nuclei dominate energy output$^8$. The IR spectra of ULIRGs display a wealth of features (Figure 1Supp), including a dozen PAH emission features$^{9,30}$, strong silicate absorption at 10\micron\ and 18\micron, an absorption complex at 6$-$8\micron\ due to water ice and hydrocarbons$^{31}$, as well as fine-structure atomic lines and rotational lines of \molh. The measurements of features essential for the analysis of this Letter are described below and are listed in Tables 1 and 2 for the 48 ULIRGs from the 1 Jy sample$^{8,32}$ (hereafter the main sample). In addition, I use a more heterogeneous sample of 98 ULIRGs with silicate strength measurements$^{10,11}$ (hereafter the auxiliary sample), of which 68 objects have \molh\ measurements in the literature$^3$. Twelve objects overlap between the main and the auxiliary samples.

Among the comparison sample of SINGS galaxies$^{4,12,30}$, I use nuclear pointings of 22 star-forming galaxies, as well as 80 extra-nuclear pointings of star-forming galaxies and galaxies with active nuclei. In the latter case, the active nucleus is sufficiently far away that it does not affect the processes captured by the extra-nuclear pointings. Using these spectra, I obtain the flux ratios of emission features shown with ellipses in Figures 2a,b and 3a,b,d. The distribution of these ratios is insensitive to whether only nuclear pointings or only extra-nuclear pointings or both are included. The \molh/F(IR) ratios shown with ellipses in Figure 3c are obtained from nuclear pointings for star-forming galaxies, for which the IR fluxes are calculated within the apertures of the spectroscopic observations$^4$. The comparison PAH/F(IR) ratios shown with dotted ellipses in Figure 2c,d are obtained using stacked template spectra$^{30}$ of all SINGS galaxies, without excluding those with active nuclei. Because of the uncertainties in the stacking procedure and because the set of galaxies included in this comparison is different from that used in other figures, the PAH/F(IR) ratios for galaxies should be regarded as an estimate. 

The apparent strength of silicate absorption is $S[9.7\micron]=-\ln(f_{\rm obs}[9.7\micron]/f_{\rm cont}[9.7\micron])$, where $f_{\rm obs}$ is the observed flux density and $f_{\rm cont}$ is the guess for the local continuum extrapolated from pivot wavelengths outside the absorption feature. Defined this way, negative values of $S[9.7\micron]$ indicate silicate emission, while positive values indicate absorption, with $S[9.7\micron]\simeq 4$ for the most absorbed sources. To estimate $f_{\rm cont}$, I average emission in the wavelength ranges $5.3-5.6\micron$ and $13.85-14.15\micron$ and then interpolate between these points using a power-law. This method is almost identical to that used for the auxiliary sample$^{11}$; the minor difference at high optical depths is that I do not use a continuum point at 7.7\micron\ even in the cases of weak PAH emission. The strength of an absorption feature differs from the optical depth of this feature by an amount corresponding to the continuum opacity at this wavelength; for typical dust opacity curves in the literature, $\tau_{\rm dust}[9.7\micron]\simeq (1.2-1.5)\times S[9.7\micron]$. 

\begin{figure}
\epsscale{0.7}
\plotone{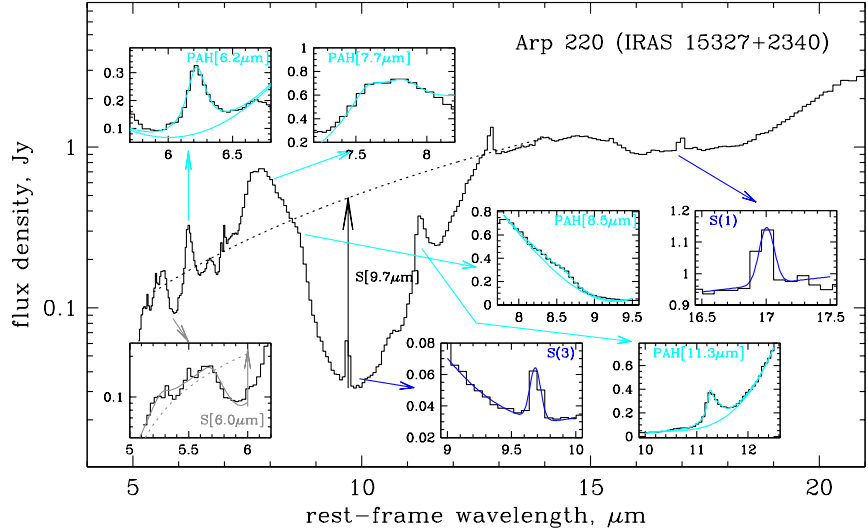}
\figcaption{{\bf Supp.} Example fits of emission and absorption features. The black dotted line shows the power-law continuum used for determination of the strength of the silicate feature (black arrow). Fit to water ice absorption is shown in grey, with the dotted line for the continuum and the arrow for the strength determination. Drude fits to PAH emission features are in light blue and Gaussian fits to \molh\ emission lines are in dark blue. \label{pic_spec}}
\end{figure}

PAH[6.2\micron, 8.5\micron, 11.3\micron] fluxes are calculated by cutting out a $\la3\micron$-wide part of the spectrum, excluding other features in that wavelength range, and then modeling this cut-out using a polynomial or a power-law continuum and Drude (damped harmonic oscillator) profiles with fixed widths$^{30}$. For PAH complexes the relative amplitudes of the components are fixed to their ratios in the template spectrum of normal star-forming galaxies$^{30}$. For example, within the 11.3\micron\ complex the amplitude ratio of the 11.23\micron\ and 11.33\micron\ components is fixed to 1.25:1. Depending on the model for the local continuum, from a constant to a cubic polynomial, the number of fit parameters varies from two to five. Because the PAH[7.7\micron] feature may be strongly affected by a number of absorption and emission features on both sides of the peak, it is fit over a narrow wavelength range (7.4$-$8.0\micron). Several measures of the luminosity of this feature are obtained by using different choices of the continuum (constant, linear, and power-law), and the slope of the spectrum at 7.5\micron\ is used as an additional luminosity measure after it is calibrated using the star-forming galaxies template$^{30}$. The median value is used as a final flux. Example fits are shown in Figure \ref{pic_spec}Supp. The quality of fits to PAH features is typically very good, and most are detected in almost all objects with high confidence. 

\begin{figure}
\epsscale{0.3}
\plotone{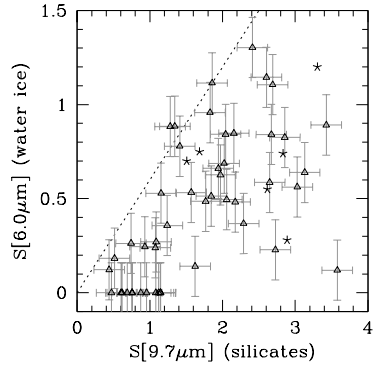}
\figcaption{{\bf Supp.} Relationship between the strengths of dust and water ice absorption features. Triangles show measurements for the main sample and stars show UGC 5101$^{43}$, IRAS 00183-7111$^{31}$, and objects from the auxiliary sample with $S[6.0\micron]$ measurements$^{44}$. Among the latter, only the four sources with wavelength coverage starting at $\lambda\le 5.6\micron$ were included (in the absence of short-wavelength coverage it is difficult to distinguish between ice absorption and PAH emission). The dotted line shows $S[6.0\micron]=0.6S[9.7\micron]$ and the error bars in the upper right corner represent estimated systematic uncertainties of the feature strength measures. \label{pic_absorption}}
\end{figure}

The strength of the 6\micron\ absorption feature is defined similarly to the silicate strength and is computed by fitting a four-parameter model to the 5$-$6\micron\ cut-out from the spectrum,
\begin{equation}
F_{\nu}=(\mbox{lin. cont.}+\mbox{PAH}[5.27\micron]+\mbox{PAH}[5.70\micron]+\mbox{PAH}[6.22\micron]))\exp\left(-S[6.0\micron]\times G(\lambda)\right),\label{eq_pah}
\end{equation}
where $G(\lambda)$ is an amplitude unity Gaussian centered at 6.0\micron\ with $\sigma=0.19\micron$. At $5\micron<\lambda<6\micron$, $G(\lambda)$ is a good approximation to the laboratory opacity of water ice$^{29}$. At longer wavelengths, this fit is not appropriate, not only because $G(\lambda)$ no longer describes water ice opacity, but also because there may be other sources of opacity contributing to the 6$-$8\micron\ complex$^{28,31}$, plus the PAH feature at 6.2\micron\ makes fitting this region very difficult. The PAH[5.27\micron] and PAH[5.70\micron] features in equation (\ref{eq_pah}) are fit with two Drude profiles using a fixed 1:2 amplitude ratio and fixed widths$^{30}$. The PAH[6.22\micron] feature is included into the fit (with a fixed 15.7:1 ratio compared to PAH[5.27\micron]), but it makes only a minor difference since only the weak extended wings of the profile contribute at wavelengths $<6\micron$. Water ice has little continuum opacity at these wavelengths, so that the measured strength of the feature is almost identical to the total optical depth, $\tau_{\rm H_2O}[6.0\micron]=1.03\times S[6.0\micron]$. $S[6.0\micron]$ is a rather difficult measurement because of the presence of multiple overlapping absorption and emission features in the same wavelength range. The relationship between silicate opacity and ice opacity is shown in Figure \ref{pic_absorption}Supp. It appears that there may be an upper limit to the amount of ice per given silicate strength, $S[6.0\micron]<0.6\times S[9.7\micron]$. Deep silicate absorption is a necessary but not sufficient condition for the presence of ice absorption. 

\molh\ line fluxes are calculated by fitting Gaussian functions plus a quadratic continuum to cut-outs from the spectrum around the lines. The widths of the Gaussians are fixed to the resolution of the spectral orders that the lines fall into$^{30}$. The rotational transitions occur in the mid-IR and are denoted as S($J$), where $J$ is the angular momentum quantum number of the final state, into which the molecule transitions from the level $J+2$. Observationally, S(1), S(2) and S(3) are the easiest to measure at low redshifts because of lack of blending with other features and relatively high signal-to-noise ratios within the spectral orders in question. Additionally, S(0), S(4) and S(7) can be measured in some cases. 

\begin{figure}
\epsscale{0.45}
\plotone{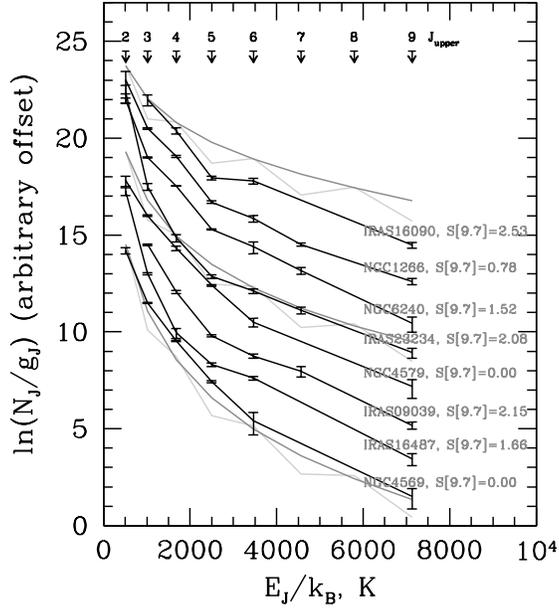}
\figcaption{{\bf Supp.} Example \molh\ excitation diagrams (black) for objects with five or more rotational lines detected, including several ULIRGs from the main sample marked 'IRAS' and several star-forming galaxies$^{4,45}$ marked 'NGC'. Objects are sorted by the S(1)/S(7) ratio, with lower ratios on top. The observer assumes an equilibrium ortho-to-para ratio of 3 to compute values on the vertical axis from line fluxes (eq. \ref{eq_population}). Excitation temperature is defined as the inverse of the slope of the curves, so high-$J$ transitions tend to have higher excitation temperatures. Arrows mark energies of the upper level with $J_{\rm upper}$ which radiatively decay into the level $J=J_{\rm upper}-2$; the corresponding line is denoted $S(J)$. Thick grey curves show excitation diagrams for models discussed in the Letter (gas emits at a range of excitation temperatures, with ${\rm d}M\propto T^{-p} {\rm d}T$). The parameter $p$ increases from 2.5 to 5.0 from top to bottom. Lighter zigzag lines show model excitation diagrams for gas with intrinsic ortho-to-para ratio of 1. Some objects show deviations from a concave excitation diagram expected in the case of the equilibrium ortho-to-para ratio. For each individual object, it is not possible to distinguish between a non-equilibrium ortho-to-para ratio (such as those seen in some nearby galaxies$^4$) and extinction on the basis of excitation diagrams alone, since the same lines (S(1) and S(3)) may suffer from both effects. The lack of a correlation between line ratios and $S[9.7\micron]$ or line deficits (parametrized by $R[J_1/J_2,J_3]$) and $S[9.7\micron]$ either for the whole sample of ULIRGs or for the well-measured subsample shown here suggests that absorption is not important. \label{pic_excitation}}
\end{figure}

Observed line fluxes $F(J)$ can be translated into level populations $N_J$:
\begin{equation}
N_{J+2}=\frac{4 \pi D_L^2 F_J}{A_{J+2\rightarrow J}(E_{J+2}-E_J)},\label{eq_population}
\end{equation}
where $D_L$ is the luminosity distance to the source, $A$ are the Einstein coefficients$^{33}$ and $E_{J+2}-E_J=hc/\lambda$ is the energy of the transition. The rotational energy levels are given by$^{34,35}$
\begin{equation}
E_J=85.35 {\rm K} \cdot k_BJ(J+1)-0.068{\rm K} \cdot k_BJ^2(J+1)^2,
\end{equation}
where $g_J$ is the degeneracy of the levels and $k_B$ is the Boltzmann constant. For an equilibrium ortho-to-para (odd $J$-to-even $J$) ratio of 3, $g_J=3(2J+1)$ for odd $J$ and $g_J=2J+1$ for even $J$. Any two or more line fluxes can be used to measure the excitation temperature as a fitting parameter according to
\begin{equation}
\frac{N_J}{g_J}\propto \exp\left(-\frac{E_J}{k_BT_{\rm exc}}\right).\label{eq_exc}
\end{equation}
In practice, gas is present at a range of excitation temperatures, from $\sim$150K to $\sim$2000K. Typical two-temperature fits to the extinction diagrams for ULIRGs$^3$, such as those in Figure \ref{pic_excitation}Supp, indicate that $\sim 10^{9}M_{\odot}$ of gas is found at $T\la 300$K and $\sim 10^6M_{\odot}$ of gas is found at $T\ga 1000$K. The massive colder component contributes very little to lines with high $J$ values, but dominates the fluxes of low-$J$ lines, while a small amount of warmer gas boosts high-$J$ fluxes. To quantify deviations from a single-temperature spectrum, for any three lines I can compute $R[J_1/J_2,J_3]$, which is the ratio of the observed flux of the S($J_1$) line to that expected from lines S($J_2$) and S($J_3$) if all three transitions were from the same excitation temperature. In the case of multi-temperature gas at an equilibrium ortho-to-para ratio and with no extinction, excitation diagrams are concave and $R$ values are less than 1 (for $J_1$ between $J_2$ and $J_3$). 

Bolometric fluxes are calculated by fitting IRAS or \spi\ broad-band data using a single-temperature black body function modified by emissivity $\propto \nu^{-1}$. The values obtained this way for IRAS data are within 10\% of those obtained using the fitting formula 
\begin{equation}
F({\rm IR})=1.8\times 10^{-11}\mbox{erg/sec/cm}^2(13.48 F_{12}+5.16 F_{25}+2.58 F_{60}+F_{100}), 
\end{equation}
where $F_{12, ...}$ are IRAS fluxes in Jy$^1$. 

The accuracy of PAH flux measurements is limited by the assumptions that were made about the shapes of the features which are fixed in my analysis. However, the results in this Letter appear to be robust to the fitting method used for calculating PAH fluxes. For example, using PAH fluxes obtained by summing up the spectrum above the local linear continuum$^{8}$ in Figures 2a,c produces similar results and the correlations with $S[9.7\micron]$ are easily detectable. As for PAH[6.2\micron] (Figure 2b), it needs to be corrected for water ice absorption (i.e., multiplied by $\exp(S[6.0\micron])$ before the correlation of PAH[11.3\micron]/PAH[6.2\micron] with $S[9.7\micron]$ becomes apparent. All conclusions remain the same if the main sample of ULIRGs is split by optical classification (LINERs vs HII regions) and each category is considered separately.

Previous studies of mid-IR \molh\ lines adopted varying views on the effects of absorption. For ULIRGs observed with ISO, \molh\ fluxes were corrected for absorption before constructing excitation diagrams$^{36}$. More recently$^3$ zero absorption correction was suggested for \molh\ in the auxiliary sample on the basis of simultaneous measurements of S(1)/S(3) and S(1)/S(2) ratios as a function of the excitation temperature (this argument is analogous to saying that $R[2/1,3]$ values in ULIRGs agree with the model value). It is largely unknown if the near-IR ro-vibrational lines in ULIRGs are co-spatial with the mid-IR rotational lines; previous studies of ULIRGs in the near-IR tend to correct the entire spectrum for extinction$^{37}$.

Several processes can lead to near-IR and mid-IR line emission from \molh$^{38}$. One possibility is excitation by UV photons of electronic levels in the molecule, which then radiatively decay into excited vibrational and rotational states. Other possibilities are collisional excitation in dense gas heated by UV or X-ray radiation, shocks or cosmic rays, or formation of hydrogen molecules in excited states on dust grains. \spi\ data by themselves cannot be used to establish the specific mechanism responsible for \molh\ excitation, because the low-energy transitions in question are almost certainly thermalized$^4$. 

Shock excitation is the explanation favored by this Letter; other possibilities appear unlikely. For example, ULIRGs presented here show no signs of an actively accreting supermassive black hole, in that no radio jets or high-ionization emission lines are seen in these objects. Besides, any component of \molh\ emission excited by a deeply buried active nucleus would be heavily extincted. The observed X-ray fluxes are too small$^{39}$ to be responsible for the observed \molh\ emission given a reasonable excitation efficiency of $10\%$ for all \molh\ lines combined$^{40}$. To estimate the contribution of cosmic rays, I assume a $10^{50}$ ergs of cosmic ray output from each supernova, and furthermore that 0.01 supernova is produced for every $M_{\odot}$ of star formation and that the star-formation rate of ULIRGs is 100$-$500 $M_{\odot}$/year$^{41}$. The luminosity in cosmic rays is then only a factor of 3$-$15 larger than the median luminosity of rotational \molh\ lines. Therefore, even if all cosmic rays are deposited directly into the interstellar medium of the galaxy (a scenario which is contradicted by some recent observations which suggest only 5\% deposition efficiency$^{42}$), somehow avoiding regions with high density and thus high obscuration, tens of per cent of their energy would need to go directly into the observed \molh\ rotational emission. The combination of these requirements makes it unlikely that cosmic rays are the dominant source of \molh\ excitation. 

{\bf Table notes.} $S[9.7\micron]$ and $S[6.0\micron]$ are the apparent strengths of the silicate and water absorption features, respectively. The small contribution of PAH and line emission to the wavelength ranges used in the calculation of $f_{\rm obs}$ and $f_{\rm cont}$ results in a systematic uncertainty in $S[9.7\micron]$ which I take to be 0.21. This value is calculated from the variance of $S[9.7\micron]$ measured in off-nuclear spectra of the comparison galaxies (when it is expected to be 0), while the statistical error is very small by comparison. I estimate the systematic uncertainty in $S[6.0\micron]$ to be 0.16 by varying the functional form of the continuum (power law instead of linear) and by computing the variance of measured $S[6.0\micron]$ for objects in which the $S[6.0\micron]=0$ fit is statistically acceptable. The nominal statistical uncertainty in the fit parameter is much smaller in most cases. Bolometric luminosities are calculated by fitting modified black-body functions (see text). Uncertainty is taken to be 10\%. PAH luminosities ($L$) and their uncertainties ($\sigma$) are in Table 1. Measurements below 2$\sigma$ are listed as 0.0. The uncertainties in PAH fluxes are calculated by adding in quadrature an estimate of the systematic error (obtained by varying the functional form of the continuum) and statistical error (derived from the standard deviation of the PAH amplitude in the linear fit). \molh\ line luminosities ($L$) and uncertainties ($\sigma$) are in Table 2. For \molh\ fluxes, I use the standard deviation of the line amplitude in the linear fit to derive flux uncertainty. Fluxes are converted to luminosities using redshifts$^{8}$ and an $h=0.7$, $\Omega_m=0.3$, $\Omega_{\Lambda}=0.7$ cosmology.

\clearpage

\clearpage
\begin{deluxetable}{cccccccccccc}
\rotate
\tabletypesize{\tiny}
\tablecaption{Absorption strengths and PAH luminosities in ULIRGs\label{tab_measures1}}
\setlength{\tabcolsep}{0.06in}
\tablewidth{0pt}
\tablehead{Object ID & $S[9.7\micron]$ & $S[6.0\micron]$ & $L$[IR] & $L$[6.2\micron] & $\sigma$[6.2\micron] & $L$[7.7\micron] & $\sigma$[7.7\micron] & $L$[8.5\micron] & $\sigma$[8.5\micron] & $L$[11.3\micron] & $\sigma$[11.3\micron]\\
& & & $10^{45}$ erg/s & $10^{42}$ erg/s & $10^{42}$ erg/s & $10^{42}$ erg/s & $10^{42}$ erg/s & $10^{42}$ erg/s & $10^{42}$ erg/s & $10^{42}$ erg/s & $10^{42}$ erg/s}
\startdata
IRAS F$00091-0738$ & 2.86 & 0.83 & 7.64 & 6.9 & 1.4 & 47.1 & 5.4 & 0.0 & 1.8 & 5.2 & 0.2 \\ 
IRAS F$00188-0856$ & 2.60 & 1.14 & 9.66 & 10.7 & 1.0 & 69.4 & 4.3 & 15.5 & 2.0 & 7.8 & 0.3 \\ 
IRAS F$00456-2904$ & 0.74 & 0.26 & 6.77 & 26.9 & 1.6 & 98.1 & 3.9 & 29.7 & 1.9 & 16.4 & 0.4 \\ 
IRAS F$00482-2721$ & 1.84 & 0.51 & 4.50 & 4.9 & 0.8 & 14.7 & 3.9 & 0.0 & 2.0 & 2.5 & 0.2 \\ 
IRAS F$01004-2237$ & 0.44 & 0.12 & 7.71 & 6.6 & 1.4 & 69.0 & 4.7 & 0.0 & 2.3 & 3.5 & 0.4 \\ 
IRAS F$01166-0844$ & 2.73 & 0.23 & 4.98 & 2.2 & 1.1 & 22.8 & 3.8 & 0.0 & 1.7 & 2.8 & 0.3 \\ 
IRAS F$01298-0744$ & 3.43 & 0.89 & 9.50 & 4.2 & 1.4 & 88.1 & 6.7 & 0.0 & 3.3 & 5.1 & 0.4 \\ 
IRAS F$01569-2939$ & 2.17 & 0.48 & 7.25 & 6.6 & 2.1 & 43.4 & 5.4 & 0.0 & 3.2 & 6.5 & 0.5 \\ 
IRAS F$02411+0353$ & 0.62 & 0.00 & 6.71 & 34.7 & 2.5 & 142.5 & 6.3 & 35.1 & 2.9 & 19.5 & 0.5 \\ 
IRAS F$03250+1606$ & 1.15 & 0.00 & 5.14 & 18.8 & 1.2 & 64.2 & 4.7 & 17.8 & 2.1 & 10.1 & 0.3 \\ 
IRAS F$04103-2838$ & 0.51 & 0.18 & 6.28 & 20.3 & 1.2 & 83.6 & 5.4 & 22.2 & 2.5 & 14.0 & 0.3 \\ 
IRAS F$08572+3915$ & 3.58 & 0.12 & 5.12 & 0.0 & 1.4 & 0.0 & 3.6 & 0.0 & 1.7 & 0.0 & 0.4 \\ 
IRAS F$09039+0503$ & 1.83 & 0.96 & 5.20 & 7.2 & 0.9 & 28.2 & 4.1 & 0.0 & 1.9 & 4.9 & 0.3 \\ 
IRAS F$09116+0334$ & 0.76 & 0.00 & 5.76 & 22.5 & 1.6 & 98.5 & 3.9 & 28.0 & 2.6 & 16.3 & 0.5 \\ 
IRAS F$09539+0857$ & 3.03 & 0.56 & 3.77 & 6.4 & 1.0 & 31.7 & 3.9 & 0.0 & 2.1 & 3.7 & 0.2 \\ 
IRAS F$10190+1322$ & 0.76 & 0.00 & 4.49 & 21.2 & 1.4 & 81.0 & 3.8 & 24.9 & 1.1 & 14.0 & 0.3 \\ 
IRAS F$10378+1108$ & 2.41 & 1.30 & 8.88 & 4.2 & 1.3 & 47.0 & 4.0 & 0.0 & 2.4 & 4.5 & 0.3 \\ 
IRAS F$10485-1447$ & 2.67 & 0.84 & 6.79 & 8.3 & 1.3 & 38.8 & 3.8 & 7.3 & 2.1 & 3.9 & 0.2 \\ 
IRAS F$10494+4424$ & 1.24 & 0.36 & 6.61 & 15.7 & 1.2 & 65.4 & 4.1 & 16.5 & 1.3 & 6.9 & 0.2 \\ 
IRAS F$11095-0238$ & 3.13 & 0.64 & 7.50 & 4.4 & 1.0 & 42.4 & 4.7 & 0.0 & 1.7 & 4.4 & 0.3 \\ 
IRAS F$11130-2659$ & 2.65 & 0.59 & 5.10 & 5.3 & 1.4 & 38.0 & 3.9 & 0.0 & 2.3 & 3.4 & 0.2 \\ 
IRAS F$11387+4116$ & 0.69 & 0.00 & 5.33 & 14.1 & 1.3 & 51.9 & 4.3 & 14.7 & 2.7 & 9.3 & 0.3 \\ 
IRAS F$11506+1331$ & 2.06 & 0.50 & 9.19 & 21.8 & 1.5 & 86.2 & 4.5 & 19.9 & 2.0 & 8.0 & 0.3 \\ 
IRAS F$12112+0305$ & 1.34 & 0.88 & 9.09 & 14.1 & 1.1 & 55.9 & 4.5 & 12.2 & 1.0 & 7.2 & 0.3 \\ 
IRAS F$12127-1412$ & 2.29 & 0.37 & 5.87 & 3.3 & 1.0 & 38.7 & 4.8 & 0.0 & 2.1 & 3.0 & 0.3 \\ 
IRAS F$12359-0725$ & 1.97 & 0.63 & 5.43 & 7.3 & 1.2 & 31.6 & 4.9 & 6.6 & 2.3 & 3.4 & 0.2 \\ 
IRAS F$13335-2612$ & 0.93 & 0.24 & 5.02 & 17.2 & 0.9 & 64.5 & 3.8 & 17.4 & 2.0 & 11.0 & 0.3 \\ 
IRAS F$13509+0442$ & 1.14 & 0.00 & 6.95 & 18.6 & 2.1 & 81.4 & 6.6 & 23.1 & 2.3 & 10.3 & 0.3 \\ 
IRAS F$13539+2920$ & 1.09 & 0.27 & 4.79 & 19.5 & 1.1 & 75.8 & 4.1 & 18.2 & 1.7 & 9.4 & 0.3 \\ 
IRAS F$14060+2919$ & 0.60 & 0.00 & 5.03 & 23.3 & 1.7 & 98.9 & 4.2 & 27.1 & 2.4 & 13.9 & 0.4 \\ 
IRAS F$14252-1550$ & 0.88 & 0.00 & 6.24 & 9.9 & 1.4 & 43.9 & 4.0 & 10.3 & 2.9 & 5.5 & 0.4 \\ 
IRAS F$14348-1447$ & 1.86 & 1.12 & 9.46 & 14.6 & 1.2 & 52.7 & 4.2 & 8.5 & 1.0 & 8.2 & 0.3 \\ 
IRAS F$15206+3342$ & 0.47 & 0.00 & 6.41 & 29.7 & 1.9 & 109.8 & 5.5 & 26.4 & 3.0 & 17.4 & 0.5 \\ 
IRAS F$15225+2350$ & 2.02 & 0.69 & 5.70 & 12.3 & 1.6 & 57.6 & 4.6 & 10.0 & 2.4 & 6.6 & 0.3 \\ 
IRAS F$15327+2340$ & 2.69 & 1.11 & 6.17 & 3.7 & 0.7 & 19.2 & 3.6 & 1.9 & 0.3 & 2.0 & 0.2 \\ 
IRAS F$16090-0139$ & 2.16 & 0.85 & 14.59 & 19.7 & 1.7 & 103.1 & 5.9 & 0.0 & 3.1 & 14.4 & 0.5 \\ 
IRAS F$16468+5200$ & 2.04 & 0.84 & 5.04 & 10.6 & 1.5 & 35.8 & 4.9 & 0.0 & 3.0 & 4.1 & 0.3 \\ 
IRAS F$16474+3430$ & 1.08 & 0.24 & 6.06 & 22.8 & 1.3 & 100.4 & 4.9 & 25.0 & 2.2 & 12.6 & 0.3 \\ 
IRAS F$16487+5447$ & 1.41 & 0.78 & 6.44 & 7.5 & 1.1 & 32.2 & 4.4 & 6.9 & 1.5 & 5.2 & 0.2 \\ 
IRAS F$17028+5817$ & 1.15 & 0.53 & 6.30 & 10.5 & 1.1 & 51.9 & 3.9 & 14.3 & 1.5 & 5.8 & 0.2 \\ 
IRAS F$17044+6720$ & 1.62 & 0.14 & 5.70 & 7.0 & 1.2 & 46.6 & 5.2 & 9.4 & 2.4 & 4.2 & 0.4 \\ 
IRAS F$20414-1651$ & 1.57 & 0.53 & 6.68 & 7.7 & 0.9 & 33.5 & 3.7 & 8.1 & 0.9 & 3.9 & 0.2 \\ 
IRAS F$21208-0519$ & 0.96 & 0.00 & 4.49 & 11.8 & 1.2 & 49.3 & 3.9 & 13.9 & 2.0 & 6.2 & 0.2 \\ 
IRAS F$21329-2346$ & 1.94 & 0.66 & 5.73 & 8.8 & 0.9 & 36.0 & 4.4 & 5.7 & 1.8 & 3.5 & 0.2 \\ 
IRAS F$22206-2715$ & 1.28 & 0.88 & 6.86 & 10.9 & 1.3 & 40.0 & 4.9 & 9.1 & 2.3 & 6.0 & 0.3 \\ 
IRAS F$22491-1808$ & 1.12 & 0.00 & 6.12 & 10.3 & 1.0 & 40.6 & 3.7 & 9.4 & 0.9 & 6.2 & 0.2 \\ 
IRAS F$23234+0946$ & 1.77 & 0.49 & 5.68 & 8.3 & 1.0 & 35.3 & 3.9 & 7.7 & 1.9 & 4.7 & 0.2 \\ 
IRAS F$23327+2913$ & 1.08 & 0.00 & 5.27 & 4.6 & 1.0 & 20.8 & 3.6 & 6.3 & 1.4 & 3.4 & 0.2 \\ 
\enddata
\end{deluxetable}

\clearpage
\begin{deluxetable}{ccccccccccc}
\rotate
\tabletypesize{\tiny}
\tablecaption{\molh\ luminosities in ULIRGs\label{tab_measures2}}
\tablewidth{0pt}
\setlength{\tabcolsep}{0.06in}
\tablehead{Object ID & $L$[S(1)] & $\sigma$[S(1)] & $L$[S(2)] & $\sigma$[S(2)] & $L$[S(3)] & $\sigma$[S(3)] & $L$[S(4)] & $\sigma$[S(4)] & $L$[S(7)] & $\sigma$[S(7)]\\
& $10^{40}$ erg/s & $10^{40}$ erg/s & $10^{40}$ erg/s & $10^{40}$ erg/s & $10^{40}$ erg/s & $10^{40}$ erg/s & $10^{40}$ erg/s & $10^{40}$ erg/s & $10^{40}$ erg/s & $10^{40}$ erg/s}
\startdata
IRAS F$00091-0738$ & 47.2 & 3.6 & 0.0 & 5.2 & 26.3 & 2.9 & 0.0 & 12.9 & 17.5 & 6.2 \\ 
IRAS F$00188-0856$ & 76.2 & 11.4 & 17.2 & 6.8 & 23.5 & 4.3 & 0.0 & 17.3 & 0.0 & 16.3 \\ 
IRAS F$00456-2904$ & 33.4 & 2.3 & 20.1 & 3.0 & 39.3 & 3.9 & 0.0 & 19.3 & 25.3 & 10.5 \\ 
IRAS F$00482-2721$ & 0.0 & 2.9 & 0.0 & 4.5 & 0.0 & 4.5 & 0.0 & 9.4 & 0.0 & 6.0 \\ 
IRAS F$01004-2237$ & 47.6 & 11.7 & 36.7 & 12.0 & 48.6 & 8.9 & 0.0 & 30.1 & 0.0 & 9.8 \\ 
IRAS F$01166-0844$ & 23.7 & 5.2 & 11.8 & 3.6 & 0.0 & 3.6 & 11.9 & 5.3 & 0.0 & 9.5 \\ 
IRAS F$01298-0744$ & 60.7 & 16.5 & 13.3 & 4.4 & 13.9 & 2.6 & 48.7 & 19.0 & 0.0 & 11.3 \\ 
IRAS F$01569-2939$ & 139.9 & 8.8 & 28.0 & 8.7 & 83.8 & 10.8 & 0.0 & 14.6 & 0.0 & 29.8 \\ 
IRAS F$02411+0353$ & 116.3 & 19.6 & 0.0 & 7.4 & 33.6 & 7.0 & 0.0 & 29.5 & 0.0 & 26.9 \\ 
IRAS F$03250+1606$ & 0.0 & 7.0 & 12.9 & 1.1 & 24.1 & 6.6 & 0.0 & 20.5 & 0.0 & 14.7 \\ 
IRAS F$04103-2838$ & 63.2 & 10.9 & 0.0 & 8.7 & 11.1 & 1.6 & 0.0 & 21.1 & 33.8 & 9.5 \\ 
IRAS F$08572+3915$ & 0.0 & 7.3 & 24.5 & 6.9 & 9.5 & 2.7 & 0.0 & 19.0 & 0.0 & 11.0 \\ 
IRAS F$09039+0503$ & 149.2 & 10.0 & 48.5 & 2.8 & 76.1 & 3.1 & 21.8 & 8.6 & 44.8 & 6.9 \\ 
IRAS F$09116+0334$ & 0.0 & 15.8 & 0.0 & 22.2 & 11.9 & 2.8 & 0.0 & 21.1 & 0.0 & 20.3 \\ 
IRAS F$09539+0857$ & 55.9 & 11.4 & 10.8 & 3.1 & 20.2 & 3.4 & 16.2 & 4.3 & 28.0 & 10.1 \\ 
IRAS F$10190+1322$ & 52.8 & 3.6 & 7.9 & 2.1 & 21.1 & 3.6 & 0.0 & 23.0 & 22.2 & 6.3 \\ 
IRAS F$10378+1108$ & 57.3 & 17.8 & 23.5 & 4.7 & 42.7 & 6.4 & 26.9 & 8.0 & 158.9 & 23.9 \\ 
IRAS F$10485-1447$ & 30.4 & 11.7 & 14.2 & 5.6 & 13.6 & 3.9 & 0.0 & 10.2 & 38.2 & 10.8 \\ 
IRAS F$10494+4424$ & 44.3 & 3.3 & 10.2 & 3.6 & 16.6 & 3.2 & 0.0 & 18.1 & 0.0 & 10.1 \\ 
IRAS F$11095-0238$ & 95.2 & 6.9 & 29.1 & 1.7 & 48.2 & 2.7 & 0.0 & 15.1 & 36.3 & 5.5 \\ 
IRAS F$11130-2659$ & 35.9 & 6.8 & 0.0 & 4.9 & 13.2 & 4.5 & 0.0 & 7.4 & 0.0 & 17.8 \\ 
IRAS F$11387+4116$ & 65.6 & 11.5 & 14.2 & 3.6 & 0.0 & 4.4 & 0.0 & 18.7 & 0.0 & 15.2 \\ 
IRAS F$11506+1331$ & 22.8 & 8.9 & 19.1 & 5.1 & 22.4 & 2.6 & 0.0 & 25.7 & 18.6 & 6.8 \\ 
IRAS F$12112+0305$ & 72.2 & 10.0 & 20.4 & 3.4 & 34.0 & 4.7 & 0.0 & 19.2 & 0.0 & 23.1 \\ 
IRAS F$12127-1412$ & 0.0 & 10.6 & 16.4 & 3.4 & 0.0 & 7.4 & 0.0 & 8.2 & 0.0 & 17.3 \\ 
IRAS F$12359-0725$ & 52.0 & 13.8 & 0.0 & 8.4 & 12.3 & 5.1 & 0.0 & 9.6 & 0.0 & 16.2 \\ 
IRAS F$13335-2612$ & 110.6 & 8.1 & 22.8 & 2.7 & 46.9 & 3.8 & 0.0 & 19.0 & 0.0 & 11.4 \\ 
IRAS F$13509+0442$ & 46.6 & 11.9 & 12.3 & 5.0 & 38.7 & 9.4 & 0.0 & 22.0 & 0.0 & 22.9 \\ 
IRAS F$13539+2920$ & 120.9 & 3.5 & 25.6 & 4.9 & 42.6 & 2.4 & 0.0 & 21.3 & 0.0 & 12.2 \\ 
IRAS F$14060+2919$ & 43.8 & 7.7 & 0.0 & 5.9 & 24.3 & 4.2 & 0.0 & 30.6 & 0.0 & 21.1 \\ 
IRAS F$14252-1550$ & 0.0 & 10.5 & 0.0 & 12.0 & 26.9 & 6.0 & 0.0 & 14.4 & 0.0 & 17.3 \\ 
IRAS F$14348-1447$ & 109.3 & 21.1 & 32.9 & 3.0 & 50.2 & 4.2 & 0.0 & 20.2 & 52.3 & 14.2 \\ 
IRAS F$15206+3342$ & 86.3 & 12.3 & 0.0 & 7.4 & 34.6 & 9.2 & 0.0 & 38.2 & 0.0 & 17.7 \\ 
IRAS F$15225+2350$ & 0.0 & 11.3 & 0.0 & 12.3 & 21.0 & 5.6 & 0.0 & 21.2 & 0.0 & 13.7 \\ 
IRAS F$15327+2340$ & 19.7 & 3.3 & 5.7 & 0.3 & 7.9 & 0.3 & 0.0 & 4.3 & 19.0 & 2.3 \\ 
IRAS F$16090-0139$ & 122.8 & 12.1 & 42.1 & 6.9 & 65.0 & 4.8 & 82.2 & 19.6 & 109.4 & 19.6 \\ 
IRAS F$16468+5200$ & 30.6 & 5.9 & 0.0 & 6.2 & 19.7 & 5.8 & 0.0 & 17.3 & 0.0 & 22.2 \\ 
IRAS F$16474+3430$ & 72.6 & 4.1 & 18.2 & 5.2 & 22.4 & 4.6 & 0.0 & 36.3 & 0.0 & 14.1 \\ 
IRAS F$16487+5447$ & 95.8 & 5.1 & 16.1 & 3.3 & 58.7 & 7.7 & 19.9 & 9.7 & 0.0 & 11.7 \\ 
IRAS F$17028+5817$ & 46.7 & 6.3 & 5.2 & 1.5 & 11.9 & 3.2 & 0.0 & 20.0 & 0.0 & 11.2 \\ 
IRAS F$17044+6720$ & 52.5 & 2.9 & 0.0 & 14.0 & 20.6 & 8.0 & 0.0 & 15.3 & 0.0 & 23.0 \\ 
IRAS F$20414-1651$ & 39.2 & 9.1 & 0.0 & 3.0 & 6.3 & 1.6 & 0.0 & 7.7 & 0.0 & 8.5 \\ 
IRAS F$21208-0519$ & 39.4 & 7.2 & 9.1 & 3.5 & 13.3 & 5.5 & 0.0 & 14.1 & 0.0 & 10.6 \\ 
IRAS F$21329-2346$ & 0.0 & 15.0 & 11.1 & 4.0 & 31.0 & 3.0 & 0.0 & 14.3 & 0.0 & 15.9 \\ 
IRAS F$22206-2715$ & 55.5 & 5.9 & 0.0 & 12.9 & 36.9 & 9.2 & 0.0 & 7.5 & 0.0 & 15.2 \\ 
IRAS F$22491-1808$ & 44.3 & 6.5 & 9.0 & 3.0 & 15.4 & 2.5 & 0.0 & 12.2 & 32.2 & 5.9 \\ 
IRAS F$23234+0946$ & 98.2 & 13.6 & 19.0 & 3.7 & 49.3 & 6.8 & 0.0 & 11.8 & 0.0 & 13.8 \\ 
IRAS F$23327+2913$ & 58.3 & 5.9 & 12.7 & 4.3 & 19.3 & 2.4 & 0.0 & 5.7 & 0.0 & 10.1 \\ 
\enddata
\end{deluxetable}

\end{document}